\documentclass[11pt]{article}
\usepackage{amsfonts,amsmath,amssymb,epsfig,float,here,latexsym,setspace}
\usepackage[all]{xy}
\usepackage[square, comma]{natbib}
\usepackage[english]{babel}
\usepackage{hyperref}
\usepackage[page]{appendix}

\def\qed{\unskip\nobreak\hfill$\Box$\par\addvspace{\medskipamount}}

\newcommand{\be}{\begin{equation}}
\newcommand{\ee}{\end{equation}}
\newcommand{\bea}{\begin{eqnarray}}
\newcommand{\eea}{\end{eqnarray}}
\newcommand{\beas}{\begin{eqnarray*}}
\newcommand{\eeas}{\end{eqnarray*}}

\newtheorem{theorem}{Theorem}[section]

\newtheorem{proposition}[theorem]{Proposition}

\newtheorem{remark}[theorem]{Remark}
\newtheorem{example}[theorem]{Example}
\newtheorem{examples}[theorem]{Examples}
\newtheorem{foo}[theorem]{Remarks}

\DeclareMathOperator{\argmax}{arg\,max}

\begin{document}
\title{
Dual Moments and Risk Attitudes\thanks{We are very grateful
to Loic Berger (discussant), Sebastian Ebert, Glenn Harrison, Richard Peter (discussant),
Nicolas Treich, Ilia Tsetlin, Bob Winkler, and, in particular, to Christian Gollier for many detailed comments
and suggestions and to Harris Schlesinger ($\dag$) for discussions.
We are also grateful to conference and seminar participants at the CEAR/MRIC Behavioral Insurance Workshop 
in Munich and
City University of London for their comments. 
This research was funded in part by the Netherlands Organization for
Scientific Research under grant NWO VIDI 2009 (Laeven).
Research assistance of Andrei Lalu is gratefully acknowledged.
Eeckhoudt: Catholic University of Lille, I\'ESEG School of Management, 3 Rue de la Digue, Lille 59000, France.
Laeven: University of Amsterdam, Amsterdam School of Economics, PO Box 15867, 1001 NJ Amsterdam, The Netherlands.
}
}
\author{Louis R. Eeckhoudt\\
{\footnotesize I\'ESEG School of Management}\\
{\footnotesize Catholic University of Lille}\\
{\footnotesize and CORE}\\
{\footnotesize {\tt Louis.Eeckhoudt@fucam.ac.be}}\\\and Roger J. A. Laeven\\
{\footnotesize Amsterdam School of Economics}\\
{\footnotesize University of Amsterdam, EURANDOM}\\
{\footnotesize and CentER}\\
{\footnotesize {\tt R.J.A.Laeven@uva.nl}}\\
[0.0cm]}
\date{This Version:
\today} \maketitle
\begin{abstract}
In decision under risk, the primal moments of mean and variance play a central role
to define the local index of absolute risk aversion.
In this paper, we show that in canonical non-EU models
dual moments have to be used instead of, or on par with, their primal counterparts to obtain
an equivalent index of absolute risk aversion.
\noindent
\\[4mm]\noindent\textbf{Keywords:}
Risk Premium; Expected Utility; Dual Theory; Rank-Dependent Utility;
Local Index; Absolute Risk Aversion.
\\[4mm]\noindent\textbf{JEL Classification:} D81.
\\[4mm]\noindent\textbf{OR/MS Classification:} Decision analysis: Risk.
\end{abstract}

\makeatletter
\makeatother
\maketitle

\newpage

\onehalfspacing

\section{Introduction}

In their important seminal work,
Pratt \cite{P64} and Arrow \cite{A65,A71} (henceforth, PA) show that
under expected utility (EU) the risk premium $\pi$
associated to a small risk $\tilde{\varepsilon}$ with zero mean
can be approximated by the following expression:
\begin{equation}
\pi\simeq\frac{\texttt{m}_{2}}{2}\left(-\frac{U''(w_{0})}{U'(w_{0})}\right).
\label{eq:pi}
\end{equation}
Here, $\texttt{m}_{2}$ is the second moment about the mean (i.e., the variance) of $\tilde{\varepsilon}$
while $U'(w_{0})$ and $U''(w_{0})$ are the first and second derivatives of the utility function of wealth $U$
at the initial wealth level $w_{0}$.\footnote{For ease of exposition, we assume $U$ to be twice continuously differentiable,
with positive first derivative.}
In the PA-approach,
the designation ``small'' refers to a risk that has a probability mass equal to unity but a small variance.
The PA-approximation in \eqref{eq:pi} provides a very insightful dissection of
the EU risk premium, disentangling the complex interplay between the
probability distribution of the risk, the decision-maker's risk attitude, and his initial wealth.
This well-known result has led to many developments and applications within the EU model
in many fields.

The aim of this paper is to show that a similar result can also be obtained outside EU,
in the dual theory of choice under risk (DT; Yaari \cite{Y87}) and, more generally and behaviorally more relevant,
under rank-dependent utility (RDU; Quiggin \cite{Q82}).
To achieve this, we substitute or complement the primal second moment $\texttt{m}_{2}$ by its dual counterpart,
and substitute or complement the derivatives of the utility function $U$ by the respective derivatives of the probability weighting function.\footnote{Dual moments
are sometimes referred to as mean order statistics in the statistics literature; see Section \ref{sec:dm} for further details.}$^{,}$\footnote{The RDU model encompasses both EU and DT as special cases and is at the basis of (cumulative) prospect theory (Tversky and Kahneman \cite{TK92}).
According to experimental evidence collected by Harrison and Swarthout \cite{HS16}, RDU seems to emerge even as the most important
non-EU preference model from a descriptive perspective.}
This modification enables us to develop for these two canonical non-EU models
a simple and intuitive local index of risk attitude that resembles the one in \eqref{eq:pi} under EU.
Our results allow for quite arbitrary utility and probability weighting functions
including inverse $s$-shaped functions
such as the probability weighting functions in Prelec \cite{P98} and Wu and Gonzalez \cite{WG96},
which are descriptively relevant (Abdellaoui \cite{A00}).
Thus, we allow for violations of strong risk aversion (Chew, Karni and Safra \cite{CKS87} and Ro\"ell \cite{R87})
in the sense of aversion to mean-preserving spreads
\`a la Rothschild and Stiglitz \cite{RS70} (see also Machina and Pratt \cite{MP97}).\footnote{\label{fn:glob}In a very stimulating strand of research,
Chew, Karni and Safra \cite{CKS87} and Ro\"ell \cite{R87}
have developed the ``global'' counterparts of the results presented here;
see also the more recent
Chateauneuf, Cohen and Meilijson \cite{CCM04,CCM05}
and Ryan \cite{R06}.
Surprisingly, the ``local'' approach has received no attention under DT and RDU,
except---to the best of our knowledge---for a relatively little used paper by Yaari \cite{Y86}.
Specifically, Yaari exploits a uniformly ordered local quotient of derivatives (his Definition 4)
with the aim to establish global results,
restricting attention to DT.
Yaari does not analyze the local behavior of the risk premium
nor does he make a reference to dual moments.
For global measures of risk aversion under prospect theory,
we refer to Schmidt and Zank \cite{SZ08}.}$^{,}$\footnote{The insightful Nau \cite{N03} proposes
a significant generalization of the PA-measure of local risk aversion in another direction.
He considers the case in which probabilities may be subjective,
utilities may be state-dependent, and probabilities and utilities may be inseparable,
by invoking
Yaari's \cite{Y69}
elementary definition of risk aversion of ``payoff convex'' preferences,
which agrees with the
Rothschild and Stiglitz \cite{RS70}
concept of aversion to mean-preserving spreads under EU.
}


Our paper is organized as follows.
In Section \ref{sec:dm} we 
define the second dual moment
and we use it in Section \ref{sec:DT} to develop the local index of absolute risk aversion under DT.
In Section \ref{sec:RDU} we extend our results to the RDU model.
Section \ref{sec:RL} discusses related literature and Section \ref{sec:examples} illustrates our results in examples.
In Section \ref{sec:portfolio} we present an application to portfolio choice
and we provide a conclusion in Section \ref{sec:con}.
Some supplementary material,
including some technical details to supplement Sections \ref{sec:DT} and \ref{sec:RDU},
the proof of a result in Section \ref{sec:RL},
and two illustrations to supplement Section \ref{sec:examples},
suppressed in this version to save space, is contained
in an online appendix.

\setcounter{equation}{0}

\section{The Second Dual Moment}\label{sec:dm}

The \textit{second dual moment about the mean} of an arbitrary risk $\tilde{\varepsilon}$, denoted by $\bar{\texttt{m}}_{2}$,
is defined by
\begin{equation}
\bar{\texttt{m}}_{2}:=
\mathbb{E}\left[\max\left(\tilde{\varepsilon}^{(1)},\tilde{\varepsilon}^{(2)}\right)\right]-\mathbb{E}\left[\tilde{\varepsilon}\right],
\label{eq:2nddm}
\end{equation}
where $\tilde{\varepsilon}^{(1)}$ and $\tilde{\varepsilon}^{(2)}$ are two independent copies of $\tilde{\varepsilon}$.
The second dual moment can be interpreted as the expectation of the largest order statistic:
it represents the expected best outcome among two independent draws of the risk.\footnote{The definition and interpretation
of the $2$-nd dual moment readily generalize to the $n$-th order, $n\in\mathbb{N}_{>0}$,
by considering $n$ copies.}

Our analysis will reveal that for an RDU maximizer who evaluates a small zero-mean risk,
the second dual moment stands on equal footing with the variance
as a fundamental measure of risk.
While the variance 
provides a measure of risk in the ``payoff plane'',\footnote{We refer to
Meyer \cite{M87} and Eichner and Wagener \cite{EW09} for insightful comparative statics results on the
mean-variance trade-off and its compatibility with EU.}
the second dual moment can be thought of as a measure of risk in the ``probability plane''.
Indeed, for a risk $\tilde{\varepsilon}$ with cumulative distribution function $F$,
so\footnote{Formally, our integrals with respect to functions are Riemann-Stieltjes integrals.
If the integrator is a cumulative distribution function of a discrete (or non-absolutely continuous) risk, or a function thereof,
then the Riemann-Stieltjes integral does not in general admit
an equivalent expression in the form of an ordinary Riemann integral.}
\begin{equation*}
\texttt{m}:=\mathbb{E}\left[\tilde{\varepsilon}\right]=\int x\,\mathrm{d}F(x),
\end{equation*}
we have that
\begin{equation*}
\texttt{m}_{2}=\int (x-\texttt{m})^{2}\,\mathrm{d}F(x),\qquad\mathrm{while}\qquad \bar{\texttt{m}}_{2}=\int (x-\texttt{m})\,\mathrm{d}(F(x))^{2}.
\end{equation*}
For the sake of brevity and in view of \eqref{eq:2nddm},
we shall term the second dual moment about the mean, $\bar{\texttt{m}}_{2}$, the \textit{maxiance}
by analogy to the \textit{variance}.
Our designation ``small'' in ``small zero-mean risk''
will quite naturally refer to a risk with small maxiance under DT and to a risk with both small variance
and small maxiance under RDU.

One readily verifies
that for a zero-mean risk $\tilde{\varepsilon}$,\footnote{This is easily seen from the Riemann-Stieltjes representations of the miniance and maxiance.
Indeed,
\begin{align*}
-\mathbb{E}\left[\min\left(\tilde{\varepsilon}^{(1)},\tilde{\varepsilon}^{(2)}\right)\right]
&=\int x\,\mathrm{d}\left(1-F(x)\right)^{2}\\
&=-2\int x\,\mathrm{d}F(x)+\int x\,\mathrm{d}\left(F(x)\right)^{2}
=\mathbb{E}\left[\max\left(\tilde{\varepsilon}^{(1)},\tilde{\varepsilon}^{(2)}\right)\right],
\end{align*}
where the last equality follows because $\int x\,\mathrm{d}F(x)=0$ when $\tilde{\varepsilon}$ is a zero-mean risk.}
\begin{equation*}
\mathbb{E}\left[\max\left(\tilde{\varepsilon}^{(1)},\tilde{\varepsilon}^{(2)}\right)\right]
=-\mathbb{E}\left[\min\left(\tilde{\varepsilon}^{(1)},\tilde{\varepsilon}^{(2)}\right)\right].
\end{equation*}
The \textit{miniance}---the expected \textit{worst} outcome among two independent draws---is perhaps a more natural measure of ``risk'',
but agrees with the maxiance for zero-mean risks upon a sign change.

Just like
the first and second primal moments occur under EU when the utility function is linear and quadratic,
the first and second dual moments correspond to a linear and quadratic probability weighting function under DT.
For further details on mean order statistics and their integral representations we refer to David \cite{D81}.
In the stochastic dominance literature, these expectations of order statistics and their higher-order generalizations
arise naturally in an interesting paper by Muliere and Scarsini \cite{MS89},
when defining a sequence of progressive $n$-th degree ``inverse'' stochastic dominances
by analogy to the conventional stochastic dominance sequence (see Ekern \cite{E80} and Fishburn \cite{F80}).\footnote{In a related strand of the literature, Eeckhoudt and Schlesinger \cite{ES06} (see also Eeckhoudt, Schlesinger and Tsetlin \cite{EST09})
and Eeckhoudt, Laeven and Schlesinger \cite{ELS16}
derive simple nested classes of lottery pairs to sign the $n$-th derivative of the utility function and probability weighting function, respectively.
Their approach can be seen to control the primal moments for EU and the dual moments for DT.}$^{,}$\footnote{Expressions similar
(but not identical) to dual moments also occur naturally in decision analysis applications.
For example, the expected value of information when the information will provide one of two signals
is the maximum of the two posterior expected values (e.g., payoffs or utilities)
minus the highest prior expected value. 
This generalizes to the case of $n > 2$ possible signals.
See Smith and Winkler \cite{SW06} for a related problem.}

\setcounter{equation}{0}

\section{Local Risk Aversion under the Dual Theory}\label{sec:DT}

Consider a DT decision-maker.
His evaluation of a risk $A$ with cumulative distribution function $F$ is given by
\begin{equation}
\int x\,\mathrm{d}h\left(F(x)\right),
\end{equation}
where the probability weighting (distortion) function $h:[0,1]\rightarrow [0,1]$ satisfies the usual properties
($h(0)=0$, $h(1)=1$, $h'(p)>0$).\footnote{\label{foot:dec}For ease of exposition, we assume $h$ to be twice continuously differentiable.}$^{,}$\footnote{Rather than distorting ``decumulative'' probabilities (as in Yaari \cite{Y87}), we adopt the convention to distort cumulative probabilities.
Our convention ensures that aversion to mean-preserving spreads corresponds to $h''<0$ (i.e., concavity) under DT,
just like it corresponds to $U''<0$ under EU, which facilitates the comparison.
Should we adopt the convention to distort decumulative probabilities,
the respective probability weighting function $\bar{h}(p):=1-h(1-p)$ would be convex when $h$ is concave.
}

In order to develop the local index of absolute risk aversion under DT we start from a lottery $A$ given by
the following representation:\footnote{In all figures,
values along (at the end of) the arrows represent probabilities (outcomes).}$^{,}$\footnote{Of course, we assume $0<p_{0}<1$.}

\vskip -0.5cm
\begin{figure}[H]
\begin{center}
\caption{Lottery $A$
}
\vskip 0.4cm
\includegraphics[scale=1.40,angle=0]{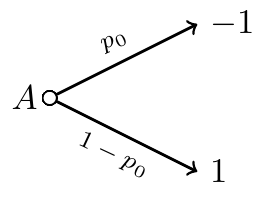}
\label{fig:DT1}
\end{center}
\end{figure}

\noindent We transform lottery $A$ into a lottery $B$ given by:\footnote{We assume $0<\varepsilon_{1}<\min\{p_{0},1-p_{0}\}$.}

\vskip -0.5cm
\begin{figure}[H]
\begin{center}
\caption{Lottery $B$
}
\vskip 0.4cm
\includegraphics[scale=1.40,angle=0]{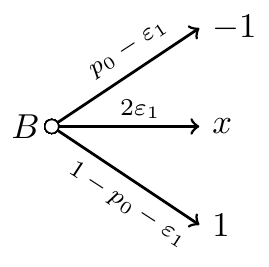}
\label{fig:DT2}
\end{center}
\end{figure}

\noindent To obtain $B$ from $A$ we subtract a probability $\varepsilon_{1}$ from the probabilities of both states of the world in $A$
without changing the outcomes
and we assign these two probabilities jointly, i.e., $2\varepsilon_{1}$, to a new intermediate state
to which we attach an outcome $x$ with $-1<x<1$.
If $x\equiv 0$, then $\mathbb{E}\left[A\right]=\mathbb{E}\left[B\right]$
and $B$ is a mean-preserving contraction of $A$.

The value of $x$ will be chosen such that the decision-maker is indifferent between $A$ and $B$.
Naturally the difference between $0$ and $x$, denoted by $\rho=0-x$,
represents the risk premium associated to the risk change from $A$ to $B$.
As we will show in Section \ref{sec:RPg}
this definition of the risk premium can be viewed as a natural generalization of the PA risk premium
to the case of risk changes with probability mass less than unity.
Depending on the shape of $h$ the risk premium $\rho$ may be positive or negative.
If (and only if) $h''(p)<0$,
the corresponding DT maximizer is averse to mean-preserving spreads,
and would universally prefer $B$ over $A$ when $x$ were $0$.\footnote{See the references in footnote \ref{fn:glob}
for global results on risk aversion under DT and RDU.}
Thus, to establish indifference between $A$ and $B$ for such a decision-maker,
$x$ has to be smaller than $0$, in which case $\rho$ is positive.

In general, for $x\equiv 0-\rho$ in $B$, indifference between $A$ and $B$ implies:
\begin{align}
&h\left(p_{0}\right)(w_{0}-1)+\left(1-h\left(p_{0}\right)\right)(w_{0}+1)\label{eq:RPDTeq}\\
&=h\left(p_{0}-\varepsilon_{1}\right)(w_{0}-1)
+\left(h\left(p_{0}+\varepsilon_{1}\right)-h\left(p_{0}-\varepsilon_{1}\right)\right)\left(w_{0}-\rho\right)
+\left(1-h\left(p_{0}+\varepsilon_{1}\right)\right)(w_{0}+1),\nonumber
\end{align}
where $w_{0}$ is the decision-maker's initial wealth level.
From \eqref{eq:RPDTeq} we obtain the explicit solution
\begin{equation}
\rho=\frac{\left(h\left(p_{0}\right)-h\left(p_{0}-\varepsilon_{1}\right)\right)-\left(h\left(p_{0}+\varepsilon_{1}\right)-h\left(p_{0}\right)\right)}
{\left(h\left(p_{0}+\varepsilon_{1}\right)-h\left(p_{0}-\varepsilon_{1}\right)\right)}.
\label{eq:RPDT}
\end{equation}

\begin{figure}[H]
\begin{center}
\caption{Mean-Preserving Spread from $B$ with $x\equiv 0$ to $A$.
}
\vskip 0.4cm
\includegraphics[scale=1.40,angle=0]{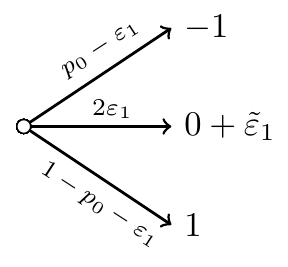}
\label{fig:DT3}
\end{center}
\end{figure}

By approximating $h\left(p_{0}\pm\varepsilon_{1}\right)$ in (\ref{eq:RPDT}) using second-order Taylor series expansions around $h(p_{0})$, 
we obtain the following approximation for the DT risk premium:
\begin{equation}
\rho\simeq\frac{\bar{\texttt{m}}_{2}}{2\texttt{Pr}}\left(-\frac{h''(p_{0})}{h'(p_{0})}\right).
\label{eq:RPDTApp}
\end{equation}
Here, $\bar{\texttt{m}}_{2}$ is the unconditional maxiance of the risk $\tilde{\varepsilon}_{1}$
that describes the mean-preserving spread from $B$ with $x\equiv 0$ to $A$.
Unconditionally, $\tilde{\varepsilon}_{1}$ takes the values $\pm 1$ each with probability $\varepsilon_{1}$.
Furthermore, $\texttt{Pr}$ 
is the total unconditional 
probability mass associated to 
$\tilde{\varepsilon}_{1}$;
see Figure \ref{fig:DT3}.
\vskip -0.5cm
\noindent
Observe
that lottery $A$ is obtained from lottery $B$ (with $x\equiv 0$) by attaching the risk $\tilde{\varepsilon}_{1}$ 
to the intermediate branch of $B$.
That is, the risk $\tilde{\varepsilon}_{1}$ is effective conditionally upon realization of the intermediate state of lottery $B$,
which occurs with probability $2\varepsilon_{1}$.
One readily verifies that, for this risk $\tilde{\varepsilon}_{1}$, we have that, unconditionally,
$\bar{\texttt{m}}_{2}=2\varepsilon_{1}^{2}$ and $\texttt{Pr}=2\varepsilon_{1}$.
We consider the unconditional maxiance of the zero-mean risk $\tilde{\varepsilon}_{1}$ to be ``small''
and compute the Taylor expansions up to the order $\varepsilon_{1}^{2}$.
Henceforth, maxiances and variances are always understood to be unconditional.


It is important to compare the result in \eqref{eq:RPDTApp}
to that obtained by PA presented in \eqref{eq:pi}.
In PA the local approximation of the risk premium is proportional to the variance,
while under DT it is proportional to the maxiance.

We note that the local approximation of the risk premium in \eqref{eq:RPDTApp}
remains valid in general,
for non-binary zero-mean risks $\tilde{\varepsilon}_{1}$ with small maxiance,
just like, as is well-known,
\eqref{eq:pi} is valid for non-binary zero-mean risks with small variance.\footnote{Detailed
derivations are suppressed to save space.
They are contained in an online appendix (available from the authors' webpages; see {\tt http://www.rogerlaeven.com}).}

\setcounter{equation}{0}

\section{Local Risk Aversion under Rank-Dependent Utility}\label{sec:RDU}

Under DT the local index arises from a risk change with small maxiance.
To deal with the RDU model, which encompasses both EU and DT as special cases,
we naturally have to consider changes in risk that are small in both variance and maxiance.
To achieve this, we start from a lottery $C$ given by:\footnote{We assume $\varepsilon_{2}>0$.}

\vskip -0.5cm
\begin{figure}[H]
\begin{center}
\caption{Lottery $C$
}
\vskip 0.4 cm
\includegraphics[scale=1.40,angle=0]{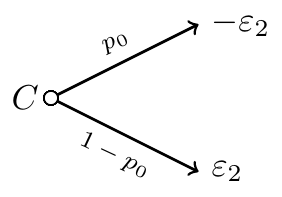}
\label{fig:RDU1}
\end{center}
\end{figure}

\noindent Similar to under DT, we transform lottery $C$ into a lottery $D$
by reducing the probabilities of both states in $C$ by a probability $\varepsilon_{1}$
and assigning the released probability $2\varepsilon_{1}$ to an intermediate state with outcome $y$,
where $-\varepsilon_{2}<y<\varepsilon_{2}$.
This yields a lottery $D$ given by:

\vskip -0.5cm
\begin{figure}[H]
\begin{center}
\caption{Lottery $D$
}
\vskip 0.4 cm
\includegraphics[scale=1.40,angle=0]{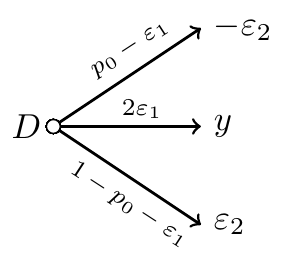}
\label{fig:RDU2}
\end{center}
\end{figure}

Of course, when $y\equiv 0$, $D$ is a mean-preserving contraction of $C$.
All RDU decision-makers that are averse to mean-preserving spreads therefore prefer $D$ over $C$ in that case.
Formally, an RDU decision-maker evaluates a lottery $A$ with cumulative distribution function $F$ by computing
\begin{equation}
\int U(x)\,\mathrm{d}h\left(F(x)\right),
\end{equation}
and is averse to mean-preserving spreads if and only if $U''<0$ and $h''<0$.\footnote{See the references in footnote \ref{fn:glob}.}

In general,
we can search for $y$ such that indifference between $C$ and $D$ occurs.
The discrepancy between the resulting $y$ and $0$ is the RDU risk premium associated to the risk change from $C$ to $D$
and its value, denoted by $\lambda=0-y$, is the solution to
\begin{align}
&
h\left(p_{0}\right)
U\left(w_{0}-\varepsilon_{2}\right)
+\left(1-h\left(p_{0}
\right)
\right)U\left(w_{0}+\varepsilon_{2}\right)\label{eq:RPRDUeq}\\
\hskip -0.1 cm =\ &h\left(p_{0}-\varepsilon_{1}\right)U\left(w_{0}-\varepsilon_{2}\right)
+\left(h\left(p_{0}+\varepsilon_{1}\right)-h\left(p_{0}-\varepsilon_{1}\right)\right)U\left(w_{0}-\lambda\right)
+\left(1-h\left(p_{0}+\varepsilon_{1}\right)\right)U\left(w_{0}+\varepsilon_{2}\right).\nonumber
\end{align}
It will be positive or negative depending on the shapes of $U$ and $h$.

Approximating the solution to \eqref{eq:RPRDUeq}
by Taylor series expansions, up to the first order in $\lambda$ around $U\left(w_{0}\right)$
and up to the second orders in $\varepsilon_{1}$ and $\varepsilon_{2}$ around $U\left(w_{0}\right)$ and $h\left(p_{0}\right)$,
we obtain the following approximation for the RDU risk premium:
\begin{equation}
\lambda\simeq\frac{\texttt{m}_{2}}{2\texttt{Pr}}\left(-\frac{U''(w_{0})}{U'(w_{0})}\right)
+\frac{\bar{\texttt{m}}_{2}}{2\texttt{Pr}}\left(-\frac{h''(p_{0})}{h'(p_{0})}\right).
\label{eq:RPRDUApp}
\end{equation}
Here, $\texttt{m}_{2}$ and $\bar{\texttt{m}}_{2}$
are the unconditional variance and maxiance of the risk $\tilde{\varepsilon}_{12}$
that dictates the mean-preserving spread from $D$ with $y\equiv 0$ to $C$.
Unconditionally, $\tilde{\varepsilon}_{12}$ takes the values $\pm\varepsilon_{2}$ each with probability $\varepsilon_{1}$.
Furthermore, $\texttt{Pr}$ 
is the total unconditional probability mass
associated to $\tilde{\varepsilon}_{12}$.\footnote{It is straightforward to verify that for $\tilde{\varepsilon}_{12}$ we have that, unconditionally,
$\texttt{m}_{2}=2\varepsilon_{1}\varepsilon_{2}^{2}$, $\bar{\texttt{m}}_{2}=2\varepsilon_{1}^{2}\varepsilon_{2}$,
and $\texttt{Pr}=2\varepsilon_{1}$.}

Comparing \eqref{eq:RPRDUApp} to \eqref{eq:pi} and \eqref{eq:RPDTApp}
reveals that the local approximation of the RDU risk premium
aggregates the (suitably scaled) EU and DT counterparts,
with the variance and maxiance featuring equally prominently.

As shown in online supplementary material,
the local approximation of the RDU risk premium in \eqref{eq:RPRDUApp}
also generalizes naturally to non-binary risks $\tilde{\varepsilon}_{12}$.

\section{Related Literature}
\label{sec:RL}

\subsection{Global Results: Comparative Risk Aversion under RDU}
\label{sec:GR}

Not only the local properties of the previous sections are valid under DT and RDU
but also the corresponding global properties,
just like in the PA-approach under the EU model (see, in particular, Theorem 1 in Pratt \cite{P64}).
In this section, we restrict attention to the RDU model.
(The DT model occurs as a special case.)
We first note that the definition of the RDU risk premium
in \eqref{eq:RPRDUeq} applies
also when $\varepsilon_{1}$ and $\varepsilon_{2}$ are ``large'',
as long as $0<\varepsilon_{1}\leq\{p_{0},1-p_{0}\}<1$ and $\varepsilon_{2}>0$
are satisfied.
We then state the following result:

\begin{proposition}
\label{th:craRDU}
Let $U_{i}$, $h_{i}$, $\lambda_{i}(p_{0},w_{0},\varepsilon_{1},\varepsilon_{2})$ be
the utility function, the probability weighting function, and the risk premium solving \eqref{eq:RPRDUeq}
for RDU decision-maker $i=1,2$.
Then the following conditions are equivalent:
\begin{itemize}
\item[(i)] $-\frac{U_{2}''(w)}{U_{2}'(w)}\geq -\frac{U_{1}''(w)}{U_{1}'(w)}$ and $-\frac{h_{2}''(p)}{h_{2}'(p)}\geq -\frac{h_{1}''(p)}{h_{1}'(p)}$
$\quad$ for all $w$ and all $p\in(0,1)$.
\item[(ii)] $\lambda_{2}(p_{0},w_{0},\varepsilon_{1},\varepsilon_{2})\geq \lambda_{1}(p_{0},w_{0},\varepsilon_{1},\varepsilon_{2})$
$\quad$ for all $0<\varepsilon_{1}\leq\{p_{0},1-p_{0}\}<1$, all $w_{0}$, and all $\varepsilon_{2}>0$.
\end{itemize}
\end{proposition}
Because the binary symmetric zero-mean risk $\tilde{\varepsilon}_{12}$ in Section \ref{sec:RDU}
induces a risk change that is a special case of a mean-preserving spread,
the implication (i)$\Rightarrow$(ii) in Proposition \ref{th:craRDU} in principle follows
from the classical results on comparative risk aversion under RDU
(Yaari \cite{Y86}, Chew, Karni and Safra \cite{CKS87}, and Ro\"ell \cite{R87}).
The reverse implication (ii)$\Rightarrow$(i) 
formalizes the connection between our local risk aversion approach and global risk aversion under RDU.

Due to the simultaneous involvement of both the utility function and the probability weighting function,
the proof of the equivalences between (i) and (ii) under RDU is more complicated than that of the analogous properties under EU (and DT).
Our proof of Proposition \ref{th:craRDU} (which is contained in online supplementary material available from the authors' webpages)
is based on the total differential of the RDU evaluation,
and the sensitivity of the risk premium
with respect to changes in payoffs.

\subsection{Relation to the Pratt-Arrow Definition of the Risk Premium}
\label{sec:RPg}

Our definition of the risk premium under RDU in \eqref{eq:RPRDUeq}
can be viewed as a natural generalization of the risk premium of Pratt \cite{P64} and Arrow \cite{A65,A71}.
To see this, first note that the PA-definition,
under which a risk is compared to a sure loss equal to the risk premium,
occurs when $p_{0}=\varepsilon_{1}=\frac{1}{2}$.\footnote{Recall that
the probability $\varepsilon_{1}$ and payoff $\pm\varepsilon_{2}$
in \eqref{eq:RPRDUeq}
can be ``large''
as long as $0<\varepsilon_{1}\leq\{p_{0},1-p_{0}\}<1$ and $\varepsilon_{2}>0$.}
%
Then, lottery $D$ becomes a sure loss of $\lambda$
the value of which is such that the decision-maker is indifferent to the risk of lottery $C$.

When $\varepsilon_{1}<\frac{1}{2}$,
our definition of the RDU risk premium provides a natural generalization
of the PA-definition.
This becomes readily apparent if we omit the common components of lotteries $D$ and $C$ with the same incremental RDU evaluation
and isolate the risk change,
which yields
\vskip -0.5cm
\begin{figure}[H]
\begin{center}
\caption{Lottery $D$ after Omitting the Components in Common with Lottery $C$.}
\vskip 0.4cm
\includegraphics[scale=1.40,angle=0]{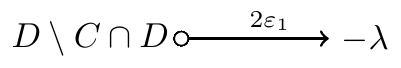}
\end{center}
\end{figure}
\noindent and
\vskip -0.5cm
\begin{figure}[H]
\begin{center}
\caption{Lottery $C$ after Omitting the Components in Common with Lottery $D$.}
\vskip 0.4cm
\includegraphics[scale=1.40,angle=0]{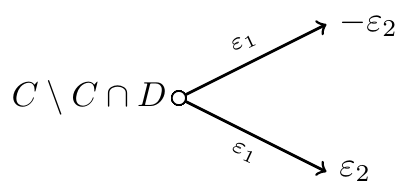}
\end{center}
\end{figure}

\noindent The value of $\lambda$ thus represents the risk premium for the risk change
induced by a risk that, unconditionally, takes the values $\pm \varepsilon_{2}$ each with probability $\varepsilon_{1}$.

When $\varepsilon_{1}<\frac{1}{2}$,
the original comparison between $C$ and $D$ is a comparison between two risky situations
as in Ross \cite{R81}, Machina and Neilson \cite{MN87}, and Pratt \cite{P90}. 
The removal of common components, however, reveals that we essentially face a PA-comparison
between a single loss and a risk with the same total probability mass,
which is now allowed to be smaller than unity.

\subsection{Related Measures of Risk}

Dual moments can be related to the Gini cofficient
named after statistician Corrado Gini and used by economists to measure
the dispersion of the income distribution of a population, summarizing its income inequality.
In risk theory,
the Gini coefficient $\mathcal{G}$ of a risk $A$
is usually defined by
\begin{equation}
\label{eq:Gini}
\mathcal{G}=\frac{\mathbb{E}\left[|A^{(1)}-A^{(2)}|\right]}{2\mathbb{E}\left[A\right]},
\end{equation}
which represents half the relative (i.e., normalized) expected absolute difference between two independent draws of the risk $A$.
One can verify that, equivalently but less well-known,
\begin{equation}
\mathcal{G}=\frac{\bar{\texttt{m}}_{2}}{\texttt{m}}.
\end{equation}
This alternative expression features the ratio of the maxiance and the first moment.

Furthermore, $n$-th degree expectations of first order statistics also appear in Cherny and Madan \cite{CM09}
as performance measures
in the context of portfolio evaluation.
In this setting, the expected maximal financial loss occurring in $n$ independent draws of a risk
is used as a measure to define an acceptability index linked to a tolerance level of stress.

\setcounter{equation}{0}

\section{Examples}\label{sec:examples}



\begin{figure}[H]
\begin{center}
\caption{Prelec's Probability Weighting Function (upper panel) and its Local Index (lower panel).
We consider $\alpha\in\{0.1,0.3,\ldots,0.9\}$.}
\vskip 0.4 cm
\includegraphics[scale=0.6,angle=0]{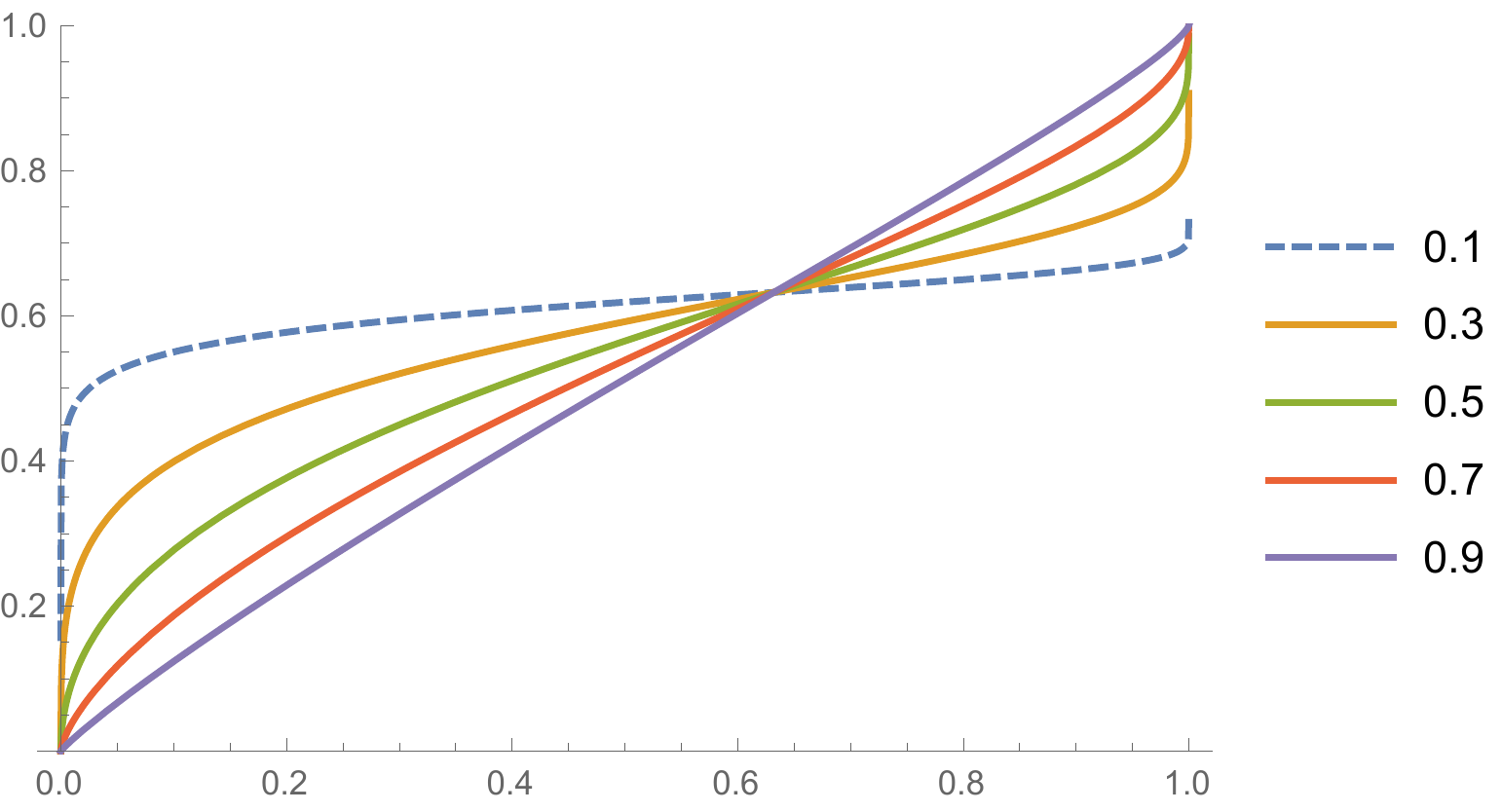}
\hskip 1.6cm
\includegraphics[scale=0.6,angle=0]{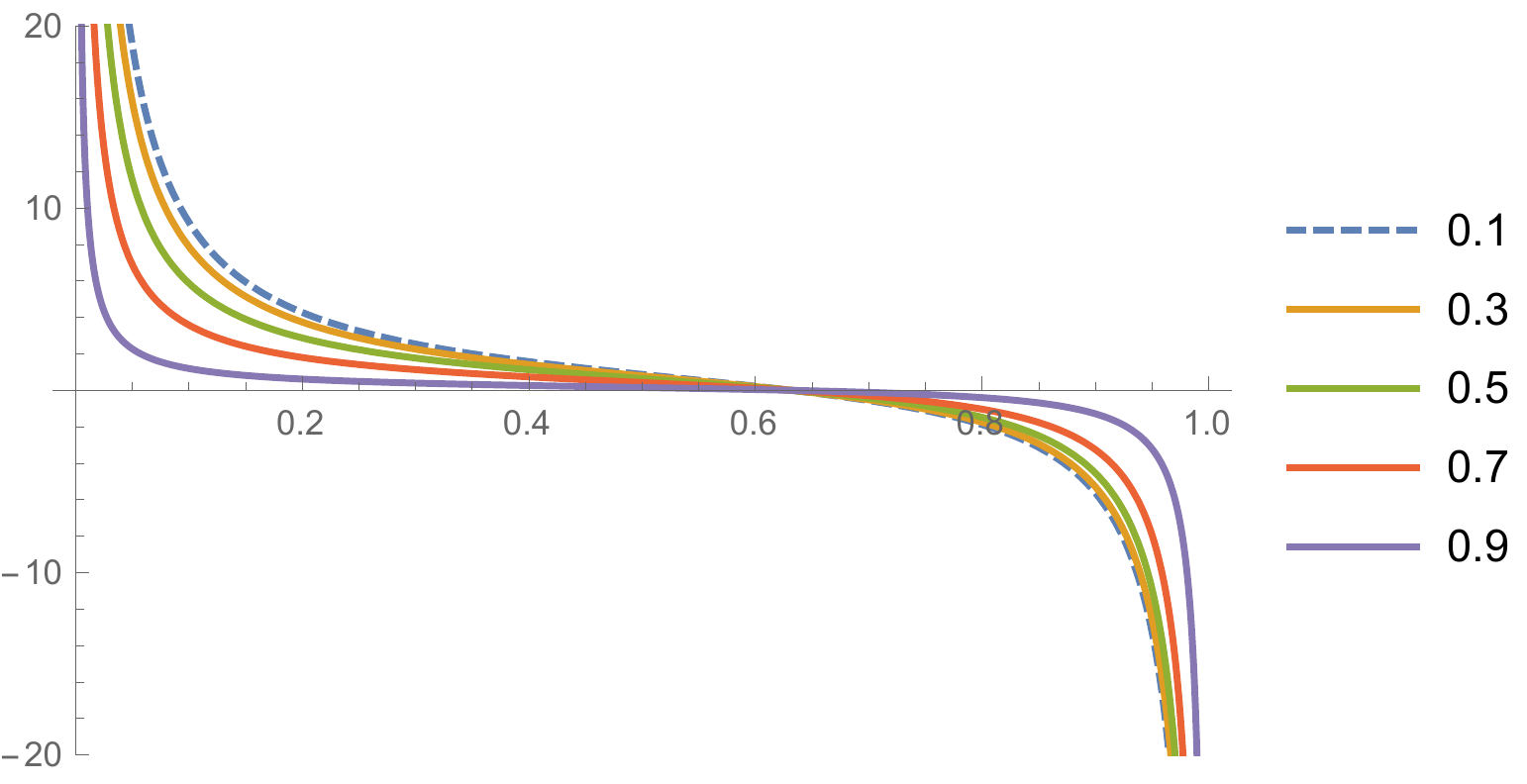}
\label{fig:Prelec}
\end{center}
\end{figure}

Owing to its local nature,
our approximation is valid and can insightfully be applied when the probability weighting function is not globally concave,
as is suggested by ample experimental evidence.
Consider the canonical probability weighting function of Prelec \cite{P98} given by\footnote{Recall our convention to distort cumulative probabilities
rather than decumulative probabilities; see footnote \ref{foot:dec}.
Prelec's original function is given by $w(p)=1-h(1-p)$.}
\begin{equation}
h(p)=1-\exp\left\{-\left(-\log \left(1-p\right)\right)^\alpha\right\},\qquad 0<\alpha<1.
\label{eq:Prel}
\end{equation}
It captures the following properties which are observed empirically:
it is regressive (first, $h(p)>p$, next $h(p)<p$), is inverse $s$-shaped (first concave, next convex), and is asymmetric (intersecting the identity probability weighting function $h(p)=p$ at $p^{*}=1-1/\exp(1)$, the inflection point).\footnote{Prelec's function is characterized axiomatically as the probability weighting function of a sign- and rank-dependent preference representation that exhibits subproportionality, 
diagonal concavity, 
and so-called \textit{compound invariance}.}
The upper panel of Figure \ref{fig:Prelec} plots this probability weighting function for
$\alpha\in\{0.1,0.3,\ldots,0.9\}$. 
(Wu and Gonzalez \cite{WG96} report estimated values of $\alpha$ between 0.03 and 0.95.)

Its local index $-\frac{h''(p)}{h'(p)}$ takes the form
\begin{equation}
-\frac{h''(p)}{h'(p)}= -\frac{1-\alpha\left(1-\left(-\log(1-p)\right)^{\alpha}\right)+\log(1-p)}{(1-p)\log(1-p)}.
\end{equation}
Figure \ref{fig:Prelec}, lower panel, plots this local index for $\alpha\in\{0.1,0.3,\ldots,0.9\}$.

\vskip -0.5cm

The inverse $s$-shape of the probability weighting function (first concave, next convex)
implies that its local index changes sign at the inflection point.
More specifically, the local index associated with Prelec's probability weighting function is decreasing (first positive, next negative) in $p$ for any $0<\alpha<1$.
This property is naturally consistent with the inverse $s$-shape property of the probability weighting function:
the inverse $s$-shape property is meant to represent a psychological phenomenon known as
\textit{diminishing sensitivity} in the probability domain (rather than the payoff domain),
under which the decision-maker is less sensitive to changes in the objective probabilities
when they move away from the reference points $0$ and $1$,
and becomes more sensitive when the objective probabilities move towards these reference points.

A decreasing local index implies in particular that $h'''>0$.
(By Pratt \cite{P64}, the sign of the derivative of the local index is the same as the sign of $\left(h''(p)\right)^{2}-h'(p)h'''(p)$.)
Inverse $s$-shaped probability weighting functions, including Prelec's canonical example, usually exhibit positive signs for the odd derivatives and alternating signs (first negative, then positive) for the even derivatives.
For a probability weighting function that is inverse $s$-shaped (first concave, then convex) and has second derivative equal to zero at the inflection point, a positive sign of the third derivative means that the function becomes increasingly concave when we move to the left of the inflection point and becomes increasingly convex when we move to the right of the inflection point.

In Figure \ref{fig:TK} in the online appendix
we also plot the local index $-\frac{h''(p)}{h'(p)}$
of the probability weighting function proposed by Tversky and Kahneman \cite{TK92} (see also Wu and Gonzalez \cite{WG96}) given by
\begin{equation}
h(p)=1-\frac{\left(1-p\right)^{\beta}}{\left(\left(1-p\right)^{\beta}+p^{\beta}\right)^{1/\beta}},\qquad 0<\beta<1,
\label{eq:TK}
\end{equation}
for values of the parameter $\beta\in\{0.55,0.65,\ldots,0.95\}$ as found in experiments
(Wu and Gonzalez \cite{WG96} report estimated values of $\beta$ between 0.57 and 0.94).
Observe the similarity between the shapes in Figure \ref{fig:Prelec} and Figure \ref{fig:TK}.

The analysis in this paper reveals that for a small risk the sign and size of the maxiance's contribution
to the RDU risk premium,
given by the second term on the right-hand side of \eqref{eq:RPRDUApp}, i.e.,
\begin{equation*}
\frac{\bar{\texttt{m}}_{2}}{2\texttt{Pr}}\left(-\frac{h''(p_{0})}{h'(p_{0})}\right),
\end{equation*}
varies with the probability level $p_{0}$,
from strongly positive to strongly negative, in tandem with the local index $-\frac{h''(p)}{h'(p)}$ to which it is proportional.

\begin{figure}[H]
\begin{center}
\caption{Surface of the RDU Risk Premium Approximation.
We consider a risk with
small variance and maxiance normalized to satisfy $\frac{\texttt{m}_{2}}{2\texttt{Pr}}=\frac{\bar{\texttt{m}}_{2}}{2\texttt{Pr}}=1$
under power utility (with $\gamma=0.5$)
and Prelec's probability weighting function (with $\alpha=0.65$).}
\vskip 0.5cm
\includegraphics[scale=0.44,angle=0]{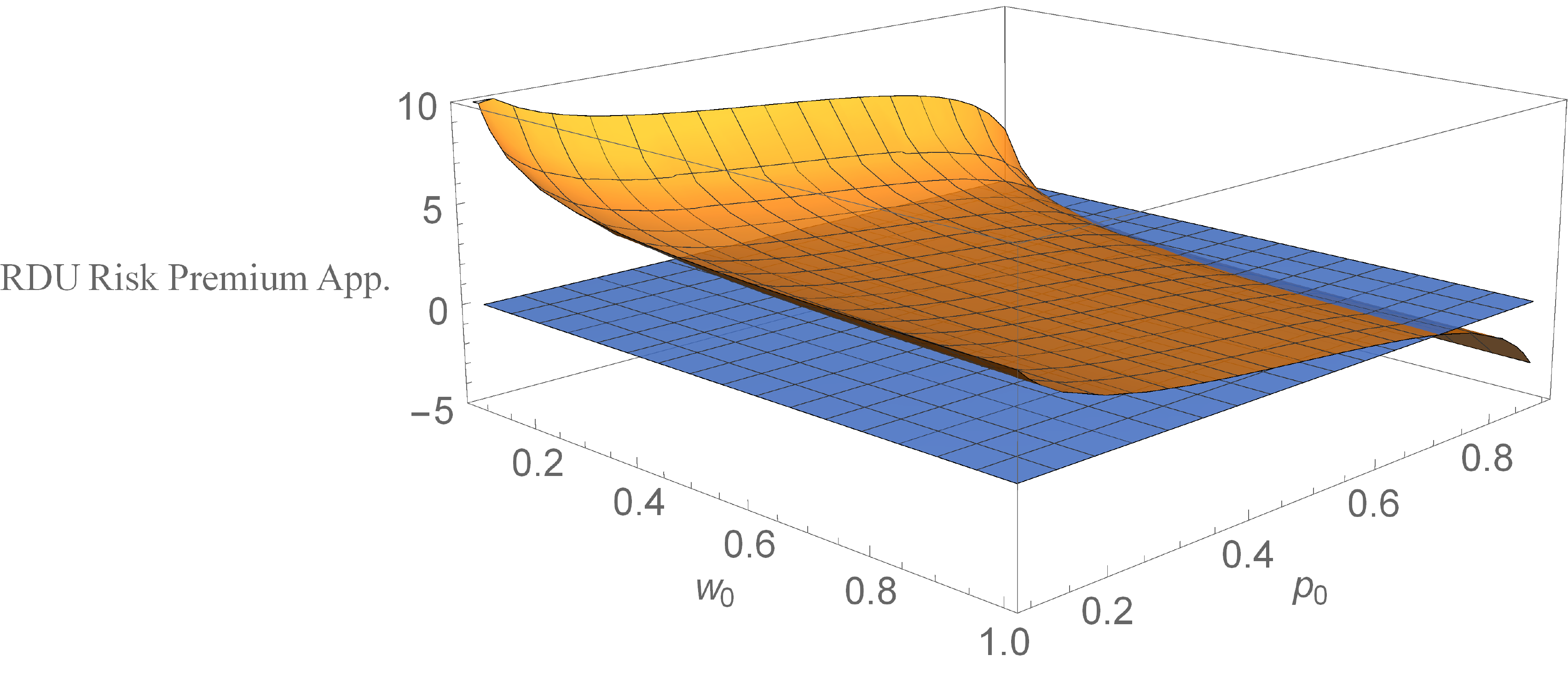}
\label{fig:RP}
\end{center}
\end{figure}

We finally plot in Figure \ref{fig:RP} our approximation to the RDU risk premium \eqref{eq:RPRDUApp}
of a risk with small variance and maxiance normalized to satisfy $\frac{\texttt{m}_{2}}{2\texttt{Pr}}=\frac{\bar{\texttt{m}}_{2}}{2\texttt{Pr}}=1$,
as a function of both the initial wealth level $w_{0}$ and the probability level $p_{0}$.
We suppose the utility function is given by the conventional power utility (note that we consider a pure rank-dependent model)
\begin{equation*}
U(x)=x^{\gamma}, 
\end{equation*}
with $\gamma=0.5$
(consistent with the gain domain in Tversky and Kahneman \cite{TK92} and with experimental evidence in Wu and Gonzalez \cite{WG96}),
and the probability weighting function is that of Prelec with parameter $\alpha=0.65$.

\vskip -0.5cm

Figure \ref{fig:RP} illustrates the interplay between the variance's and the maxiance's contributions
to the RDU risk premium \eqref{eq:RPRDUApp},
depending on the local indices $\left(-\frac{U''(w)}{U'(w)}\right)$ and $\left(-\frac{h''(p)}{h'(p)}\right)$
evaluated in the wealth and probability levels $w_{0}$ and $p_{0}$, respectively.
The orange surface represents our approximation \eqref{eq:RPRDUApp} to the RDU risk premium $\lambda$,
while the blue surface is the $\lambda=0$-plane.
To illustrate the effect of a change in variance or maxiance,
we also plot in Figure \ref{fig:RPvarmax} in the online appendix
the surface of the RDU risk premium approximation \eqref{eq:RPRDUApp}
for a small risk with ratio between the variance and maxiance equal to $3$ (upper panel)
and $1/3$ (lower) panel, instead of a ratio of $1$ as in Figure \ref{fig:RP}.

\setcounter{equation}{0}

\section{A Portfolio Application}\label{sec:portfolio}

In order to illustrate how the concepts we have developed can be used
we consider a simple portfolio problem with a safe asset,
the return of which is zero,
and a binary risky asset with returns expressed by the following representation:\footnote{We assume $0<R_{0}<R_{1}$.}

\vskip -0.5cm
\begin{figure}[H]
\begin{center}
\caption{Return Distribution of the Risky Asset
}
\vskip 0.4 cm
\includegraphics[scale=1.40,angle=0]{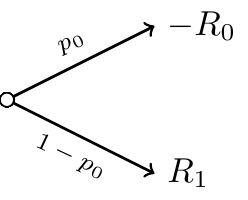}
\label{fig:R0}
\end{center}
\end{figure}

\noindent Taking $\frac{R_{1}}{R_{0}+R_{1}}>p_{0}$ makes the expected return strictly positive.

If an RDU investor has initial wealth $w_{0}$ his portfolio optimization problem
is given by
\begin{equation}
\argmax_{\alpha}\left\{h\left(p_{0}\right)U\left(w_{0}-\alpha R_{0}\right)
+\left(1-h\left(p_{0}\right)\right)U\left(w_{0}+\alpha R_{1}\right)\right\},
\label{eq:portfolio}
\end{equation}
with first-order condition (FOC) given by
\begin{equation*}
-R_{0}h\left(p_{0}\right)U'\left(w_{0}-\alpha R_{0}\right)
+R_{1}\left(1-h\left(p_{0}\right)\right)U'\left(w_{0}+\alpha R_{1}\right)\equiv 0.
\end{equation*}
It is straightforward to show that the second-order condition for a maximum is satisfied
provided $U''<0$.

Let us now pay attention to the RDU investor for whom it is optimal to choose not to invest in the risky asset,
i.e., to select $\alpha\equiv 0$.
Plugging $\alpha\equiv 0$ into the FOC we obtain the condition
\begin{equation}
h\left(p_{0}\right)\equiv\frac{R_{1}}{R_{0}+R_{1}}.
\label{eq:alpha0}
\end{equation}
Without surprise, $h\left(p_{0}\right)>p_{0}$.
This value of $h\left(p_{0}\right)$ expresses the intensity of risk aversion
that induces the choice of $\alpha\equiv 0$.

Now consider a mean-preserving contraction of the return of the risky asset given by:

\vskip -0.5cm
\begin{figure}[H]
\begin{center}
\caption{Mean-Preserving Contraction of the Risky Asset
}
\vskip 0.4 cm
\includegraphics[scale=1.40,angle=0]{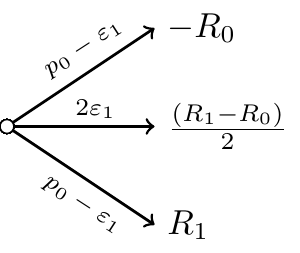}
\label{fig:R1}
\end{center}
\end{figure}

\noindent One may verify that such a mean-preserving contraction for a decision-maker who had decided not to participate in the risky asset
may induce him to select a strictly positive $\alpha$.

Hence, we raise the following question:
By how much should we reduce the intermediate return $\frac{R_{1}-R_{0}}{2}$
to induce the decision-maker to stick to the optimal $\alpha$ equal to zero?
The answer to this question is denoted by $\varsigma$.

Because we are concentrating on the situation where $\alpha\equiv 0$ is optimal,
the analysis is related only to the shape of the probability weighting function.
Indeed, the shape of $U$ that appears in the FOC through different values of $U'$
becomes irrelevant at $\alpha\equiv 0$.
The reason to concentrate on $\alpha\equiv 0$ where only the probability weighting function matters under RDU
pertains to the well-known fact that 
under EU a mean-preserving contraction of the risky return has an ambiguous effect on the optimal $\alpha$
(Gollier \cite{G95}).

It turns out that, upon invoking Taylor series expansions and after several basic manipulations,
the reduction $\varsigma$ that answers our question raised above is given by
\begin{equation}
\varsigma\simeq\frac{\bar{\texttt{m}}_{2}}{2\texttt{Pr}}\left(-\frac{h''\left(p_{0}\right)}{h'\left(p_{0}\right)}\right),
\end{equation}
where $\bar{\texttt{m}}_{2}$ is the 
maxiance of the risk
that, unconditionally, takes the values $\pm\frac{R_{0}+R_{1}}{2}$
each with probability $\varepsilon_{1}$,
and where $\texttt{Pr}$ 
is the total probability mass 
of this risk.
Again the second dual moment (instead of the primal one) appears,
jointly with the intensity of risk aversion induced by the probability weighting function.
In particular, the mean-preserving contraction is an improvement
and has made the risky asset attractive
if and only of $\varsigma$ is positive.

\section{Conclusion}\label{sec:con}

Under EU, the risk premium is approximated by an expression that multiplies
half the variance of the risk (i.e., its primal second central moment)
by the local index of absolute risk aversion.
This expression dissects the complex interplay between the risk's probability distribution,
the decision-maker's preferences and his initial wealth
that the risk premium in general depends on.
Surprisingly, a similar expression almost never appears in non-EU models. 

In this paper, we have shown that when one refers to the second dual moment---instead of, or on par with, the
primal one---one obtains quite naturally an approximation of the risk premium
in canonical non-EU models
that mimics the one obtained within the EU model.

The PA-approximation of the risk premium under EU
has induced thousands of applications
and results in many fields such as operations research, insurance, finance, and environmental economics.
So far, comparable developments 
have been witnessed to a much lesser extent outside the EU model. 
Hopefully, the new and simple expression of the approximated risk premium
may contribute to a widespread analysis and use of risk premia for non-EU.




\begin{spacing}{0.0}

\end{spacing}

\newpage

\appendix

\vskip 1cm

\begin{center}
\noindent {\Huge SUPPLEMENTARY MATERIAL\\[4mm]
(FOR ONLINE PUBLICATION)}
\end{center}

\newpage

\setcounter{equation}{0}

\section{Generalization to Non-Binary Risks}

In this supplementary material,
we first show that the local approximation for the DT risk premium in \eqref{eq:RPDTApp}
remains valid for non-binary risks with small maxiance.
Next, we prove that the RDU risk premium approximation in \eqref{eq:RPRDUApp} also remains valid
for non-binary risks with both small variance and small maxiance.
Throughout this supplement, we consider $n$-state risks
with probabilities $p_{i}$ associated to outcomes $x_{i}$, $i=1,\ldots,n$, with $n\in\mathbb{N}_{>0}$.
We order states from the lowest outcome state (designated by state number 1)
to the highest outcome state (designated by state number $n$), which means that $x_{1}\leq\cdots\leq x_{n}$.

We analyze the DT risk premium for a risk with $n\geq 2$ effective states
that have equal unconditional probability of occurrence given by $\frac{2\varepsilon_{1}}{n}$, $0<\varepsilon_{1}\leq\frac{1}{2}$.
The outcomes are, however, allowed to be the same among adjacent states;
this would 
correspond to a risk with non-equal state probabilities.
Note the generality provided by this construction.
We suppose that the risk has mean equal to zero, so $\sum_{i=1}^{n}x_{i}=0$.
One may verify that the unconditional maxiance of this $n$-state risk is given by
\begin{equation}
\bar{\texttt{m}}_{2}=\frac{4\varepsilon_{1}^{2}}{n^{2}}\sum_{i=1}^{n}\left(2i-1\right)x_{i},
\label{eq:maxgen}
\end{equation}
and that the total 
probability mass $\texttt{Pr}=2\varepsilon_{1}$.
Observe that the maxiance is of the order $\varepsilon_{1}^{2}$, i.e., $\bar{\texttt{m}}_{2}=O\left(\varepsilon_{1}^{2}\right)$.

Similar to the main text, this zero-mean risk is attached 
to the intermediate branch of lottery $B$ (with $x\equiv 0$)
to induce a mean-preserving spread.
(We normalize the outcomes of the zero-mean risk by restricting them to the interval $[-1,1]$.
This ensures that the initial ordering of outcomes in lottery $B$ is not affected
and can easily be generalized.)
The DT risk premium $\rho$ then occurs as the solution to
\begin{align}
&\left(h\left(p_{0}+\varepsilon_{1}\right)-h\left(p_{0}-\varepsilon_{1}\right)\right)\left(w_{0}-\rho\right)\nonumber\\
&=\sum_{i=1}^{n}\left(h\left(p_{0}-\varepsilon_{1}+i\frac{2\varepsilon_{1}}{n}\right)
-h\left(p_{0}-\varepsilon_{1}+\left(i-1\right)\frac{2\varepsilon_{1}}{n}\right)\right)\left(w_{0}+x_{i}\right).
\label{eq:RPDTeqgen}
\end{align}
From \eqref{eq:RPDTeqgen} we obtain the explicit solution
\begin{equation}
\rho=-\sum_{i=1}^{n}\frac{\left(h\left(p_{0}-\varepsilon_{1}+i\frac{2\varepsilon_{1}}{n}\right)
-h\left(p_{0}-\varepsilon_{1}+\left(i-1\right)\frac{2\varepsilon_{1}}{n}\right)\right)}
{h\left(p_{0}+\varepsilon_{1}\right)-h\left(p_{0}-\varepsilon_{1}\right)}x_{i}.
\label{eq:RPDTgen}
\end{equation}

By invoking Taylor series expansions around $h\left(p_{0}\right)$
up to the second order in $\varepsilon_{1}$
we obtain from \eqref{eq:RPDTgen} the following approximation for the DT risk premium:
\begin{align*}
\rho&\simeq
-\sum_{i=1}^{n}
\frac{\frac{1}{2}
\left(2i-1\right)\frac{4\varepsilon_{1}^{2}}{n^{2}}h''\left(p_{0}\right)}
{2\varepsilon_{1}h'\left(p_{0}\right)}x_{i}\\
&=\frac{\bar{\texttt{m}}_{2}}{2\texttt{Pr}}\left(-\frac{h''(p_{0})}{h'(p_{0})}\right),
\end{align*}
where the last equality follows directly from \eqref{eq:maxgen}.

Finally, turning to the risk premium under RDU,
we consider, as under DT, an $n$-state zero-mean risk
with unconditional state probabilities $\frac{2\varepsilon_{1}}{n}$,
so $\sum_{i=1}^{n}x_{i}=0$ and $\texttt{Pr}=2\varepsilon_{1}$,
now assumed to satisfy additionally that $\texttt{m}_{2}=\frac{1}{n}\sum_{i=1}^{n}x_{i}^{2}=O\left(\varepsilon_{2}^{2}\right)$
for some $\varepsilon_{2}>0$.
Upon attaching this zero-mean risk 
to the intermediate branch of lottery $D$ (with $y\equiv 0$ and assuming without losing generality that $|x_{i}|<\varepsilon_{2}$),
the RDU risk premium $\lambda$ occurs as the solution to
\begin{align}
&\left(h\left(p_{0}+\varepsilon_{1}\right)-h\left(p_{0}-\varepsilon_{1}\right)\right)U\left(w_{0}-\lambda\right)\nonumber\\
&=\sum_{i=1}^{n}\left(h\left(p_{0}-\varepsilon_{1}+i\frac{2\varepsilon_{1}}{n}\right)
-h\left(p_{0}-\varepsilon_{1}+\left(i-1\right)\frac{2\varepsilon_{1}}{n}\right)\right)U\left(w_{0}+x_{i}\right).
\label{eq:RPRDUeqgen}
\end{align}

Invoking Taylor series expansions
up to the first order in $\lambda$ around $U\left(w_{0}\right)$
and up to the second order in $x_{i}$ and $\varepsilon_{1}$ around $U\left(w_{0}\right)$ and $h\left(p_{0}\right)$,
we obtain from \eqref{eq:RPRDUeqgen}, at the leading orders, the desired approximation for the RDU risk premium:
\begin{align*}
\lambda
&\simeq-\sum_{i=1}^{n}
\frac{\frac{1}{2}\frac{2\varepsilon_{1}}{n}U''\left(w_{0}\right)}
{2\varepsilon_{1}U'\left(w_{0}\right)}x_{i}^{2}
-\sum_{i=1}^{n}
\frac{\frac{1}{2}
\left(2i-1\right)\frac{4\varepsilon_{1}^{2}}{n^{2}}h''\left(p_{0}\right)}
{2\varepsilon_{1}h'\left(p_{0}\right)}x_{i}\\
&=
\frac{\texttt{m}_{2}}{2\texttt{Pr}}\left(-\frac{U''(w_{0})}{U'(w_{0})}\right)
+\frac{\bar{\texttt{m}}_{2}}{2\texttt{Pr}}\left(-\frac{h''(p_{0})}{h'(p_{0})}\right).
\end{align*}

\newpage

\section{Proof of Proposition \ref{th:craRDU}}
First, note that (i) is equivalent to
\begin{itemize}
\item[(iv)] $U_{2}(U_{1}^{-1}(t))$ and $h_{2}(h_{1}^{-1}(u))$ are concave functions of $t$ and $u$ $\quad$ for all $t$ and all $u\in(0,1)$.
\item[(v)] $\frac{U_{2}(y)-U_{2}(x)}{U_{2}(w)-U_{2}(v)}\leq \frac{U_{1}(y)-U_{1}(x)}{U_{1}(w)-U_{1}(v)}$
and
$\frac{h_{2}(s)-h_{2}(r)}{h_{2}(q)-h_{2}(p)}\leq \frac{h_{1}(s)-h_{1}(r)}{h_{1}(q)-h_{1}(p)}$
$\quad$ for all $v<w\leq x<y$ and all $0<p<q\leq r<s<1$.
\end{itemize}
The equivalence of (i), (iv) and (v) follows trivially from the equivalence of (a), (d) and (e) in Theorem 1 of Pratt \cite{P64}
and the corresponding DT counterpart equivalences. 

Second, we will prove that (the equivalent) (i), (iv) and (v) imply (ii). 
Reconsider \eqref{eq:RPRDUeq}.
Fix (a feasible) $\varepsilon_{1}>0$ (satisfying $0<\varepsilon_{1}\leq\{p_{0},1-p_{0}\}<1$).
Note that if we let $\varepsilon_{2}\rightarrow 0$ in \eqref{eq:RPRDUeq}, then $\lambda_{i}\rightarrow 0$.
Define
\begin{align*}
V_{i}(\lambda_{i},\varepsilon_{2})=&\left(h_{i}(p_{0}+\varepsilon_{1})-h_{i}(p_{0}-\varepsilon_{1})\right)U_{i}\left(w_{0}-\lambda_{i}\right)\nonumber\\
&-\left(\left(h_{i}(p_{0})-h_{i}(p_{0}-\varepsilon_{1})\right)U_{i}(w_{0}-\varepsilon_{2})
+\left(h_{i}(p_{0}+\varepsilon_{1})-h_{i}(p_{0})\right)U_{i}(w_{0}+\varepsilon_{2})\right).
\end{align*}
We compute the total differential
$\mathrm{d}V_{i}=\frac{\partial V_{i}}{\partial \lambda_{i}}\,\mathrm{d}\lambda_{i} + \frac{\partial V_{i}}{\partial \varepsilon_{2}}\,\mathrm{d}\varepsilon_{2}.
$
It is given by
\begin{align*}
-&\left(h_{i}(p_{0}+\varepsilon_{1})-h_{i}(p_{0}-\varepsilon_{1})\right)U'_{i}\left(w_{0}-\lambda_{i}\right)\,\mathrm{d}\lambda_{i}\\
&+\left(\left(h_{i}(p_{0})-h_{i}(p_{0}-\varepsilon_{1})\right)U'_{i}(w_{0}-\varepsilon_{2})
-\left(h_{i}(p_{0}+\varepsilon_{1})-h_{i}(p_{0})\right)U'_{i}(w_{0}+\varepsilon_{2})\right)\,\mathrm{d}\varepsilon_{2}.
\end{align*}
Equating the total differential to zero yields
\begin{align}\label{eq:tdrp}
\frac{\mathrm{d}\lambda_{i}}{\mathrm{d}\varepsilon_{2}}=
\frac{h_{i}(p_{0})-h_{i}(p_{0}-\varepsilon_{1})}{h_{i}(p_{0}+\varepsilon_{1})-h_{i}(p_{0}-\varepsilon_{1})}\frac{U'_{i}(w_{0}-\varepsilon_{2})}{U'_{i}(w_{0}-\lambda_{i})}
-\frac{h_{i}(p_{0}+\varepsilon_{1})-h_{i}(p_{0})}{h_{i}(p_{0}+\varepsilon_{1})-h_{i}(p_{0}-\varepsilon_{1})}\frac{U'_{i}(w_{0}+\varepsilon_{2})}{U'_{i}(w_{0}-\lambda_{i})}.
\end{align}

From (i), as in Pratt \cite{P64}, Eqn. (20),
\begin{equation*}
\frac{U_{2}'(x)}{U_{2}'(w)}\leq \frac{U_{1}'(x)}{U_{1}'(w)},\qquad\mathrm{for}\ w<x,
\qquad\mathrm{and}\qquad
\frac{U_{2}'(x)}{U_{2}'(y)}\geq \frac{U_{1}'(x)}{U_{1}'(y)},\qquad\mathrm{for}\ x<y.
\end{equation*}
Furthermore, from (v),
\begin{equation*}
\frac{U_{2}(y)-U_{2}(x)}{U_{2}(w)-U_{2}(v)}+\frac{U_{2}(w)-U_{2}(v)}{U_{2}(w)-U_{2}(v)}
\leq \frac{U_{1}(y)-U_{1}(x)}{U_{1}(w)-U_{1}(v)}+\frac{U_{1}(w)-U_{1}(v)}{U_{1}(w)-U_{1}(v)},\qquad\mathrm{for}\ v<w\leq x<y.
\end{equation*}
Taking $w=x$ yields
\begin{equation*}
\frac{U_{2}(y)-U_{2}(v)}{U_{2}(w)-U_{2}(v)}
\leq \frac{U_{1}(y)-U_{1}(v)}{U_{1}(w)-U_{1}(v)},\qquad\mathrm{for}\ v<w<y,
\end{equation*}
hence
\begin{equation*}
\frac{U_{2}(w)-U_{2}(v)}{U_{2}(y)-U_{2}(v)}
\geq \frac{U_{1}(w)-U_{1}(v)}{U_{1}(y)-U_{1}(v)},
\quad\mathrm{and\ also}\quad
\frac{U_{2}(y)-U_{2}(w)}{U_{2}(y)-U_{2}(v)}
\leq \frac{U_{1}(y)-U_{1}(w)}{U_{1}(y)-U_{1}(v)},
\end{equation*}
for $v<w<y$.
In all inequalities in this paragraph, $U_{i}$ may be replaced by $h_{i}$, with $v,w,x$ and $y$ restricted to $(0,1)$.

Thus, from \eqref{eq:tdrp} and the inequalities above,
\begin{equation}\label{eq:tdrp2}
\frac{\mathrm{d}\lambda_{2}}{\mathrm{d}\varepsilon_{2}}\geq \frac{\mathrm{d}\lambda_{1}}{\mathrm{d}\varepsilon_{2}},
\end{equation}
hence (ii).

We have now proved that (ii) is implied by (the equivalent) (i), (iv) and (v).
We finally show that (ii) implies (i),
or rather that not (i) implies not (ii).
This goes by realizing that, by the arbitrariness of $w_{0}$, $p_{0}$,
$\varepsilon_{1}$ with $0<\varepsilon_{1}\leq\{p_{0},1-p_{0}\}<1$,
and $\varepsilon_{2}>0$,
if (i) does not hold on some interval (of $w$ or $p$),
one can always find feasible $w_{0}$, $p_{0}$, $\varepsilon_{1}$ and $\varepsilon_{2}$, such that
\eqref{eq:tdrp2}, 
hence (ii), 
hold on some interval but with the inequality signs strict and flipped.
\qed

\newpage
\section{Figures}

\vskip -0.5cm
\begin{figure}[H]
\begin{center}
\caption{Tversky-Kahneman Probability Weighting Function (upper panel) and its Local Index (lower panel).
We consider $\beta\in\{0.55,0.65,\ldots,0.95\}$.}
\vskip 0.4 cm
\includegraphics[scale=0.6,angle=0]{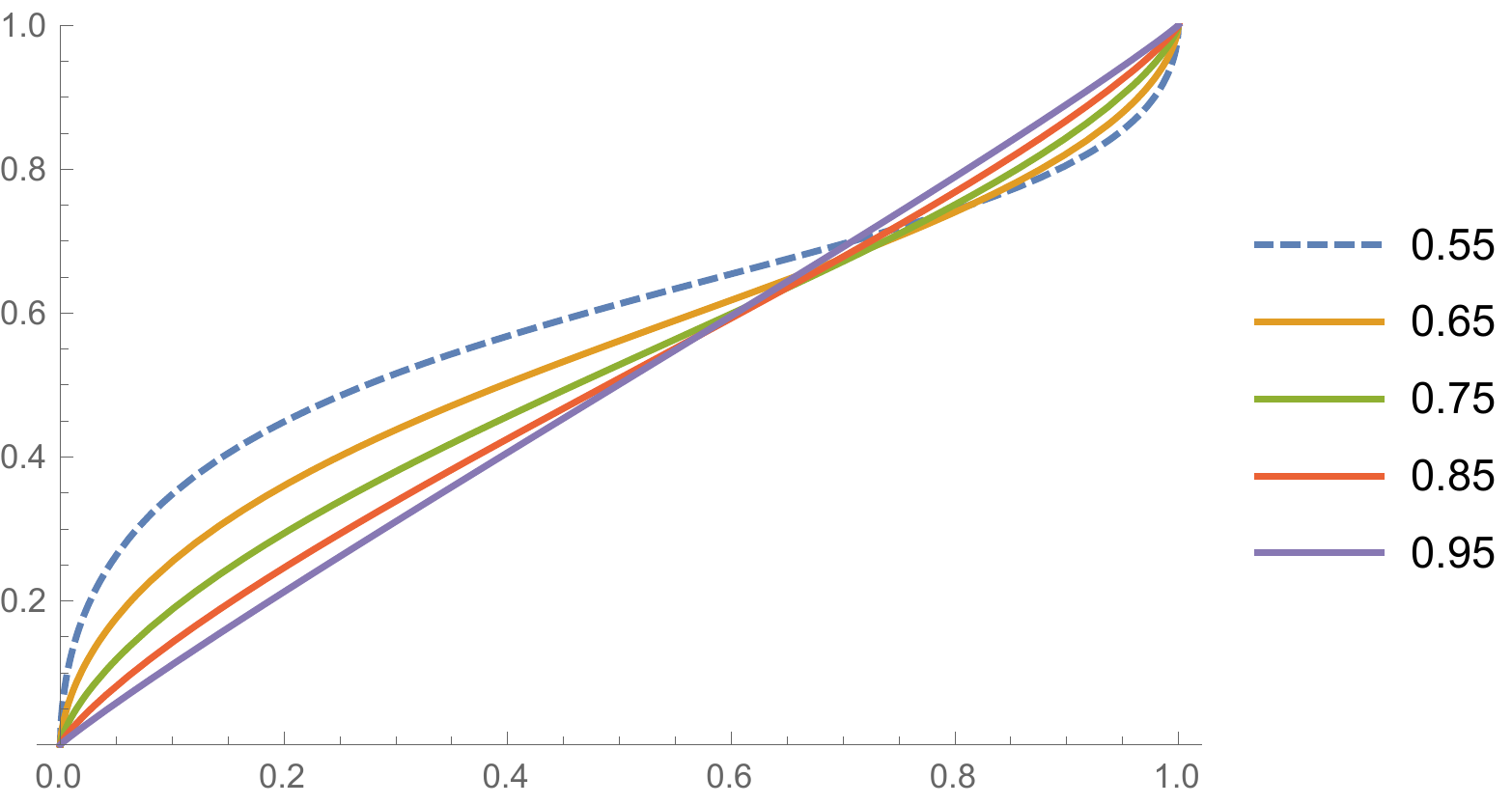}
\includegraphics[scale=0.6,angle=0]{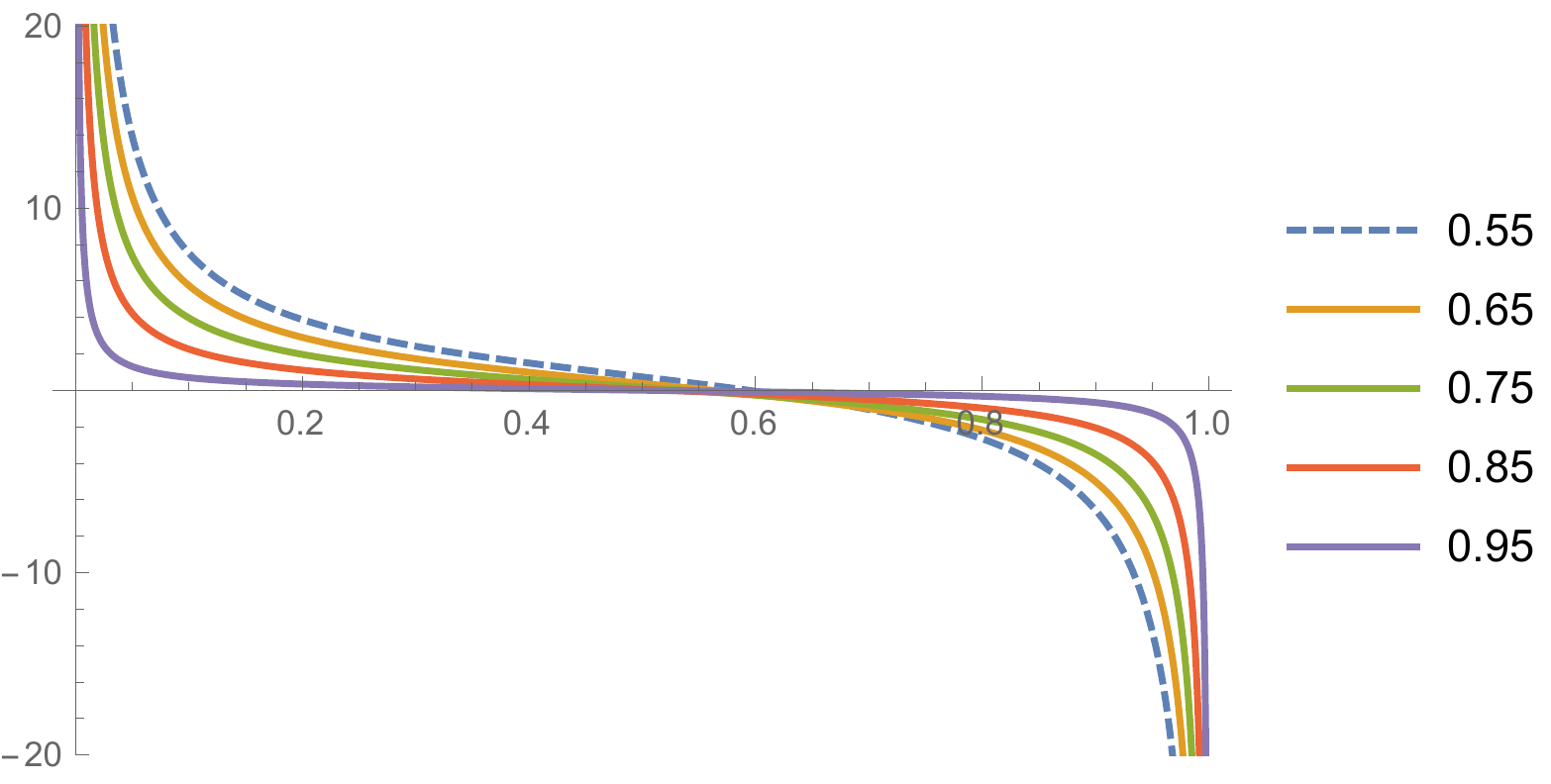}
\label{fig:TK}
\end{center}
\end{figure}

\vskip -0.5cm
\begin{figure}[H]
\begin{center}
\caption{Surface of the RDU Risk Premium Approximation.
We consider a risk with
small variance and maxiance, normalized to satisfy
$\frac{\bar{\texttt{m}}_{2}}{2\texttt{Pr}}=1$ with $\texttt{m}_{2}/\bar{\texttt{m}}_{2}=3$ (upper panel)
and $\frac{\texttt{m}_{2}}{2\texttt{Pr}}=1$ with $\texttt{m}_{2}/\bar{\texttt{m}}_{2}=1/3$ (lower panel),
under power utility (with $\gamma=0.5$)
and Prelec's probability weighting function (with $\alpha=0.65$).}
\vskip 0.5cm
\includegraphics[scale=0.43,angle=0]{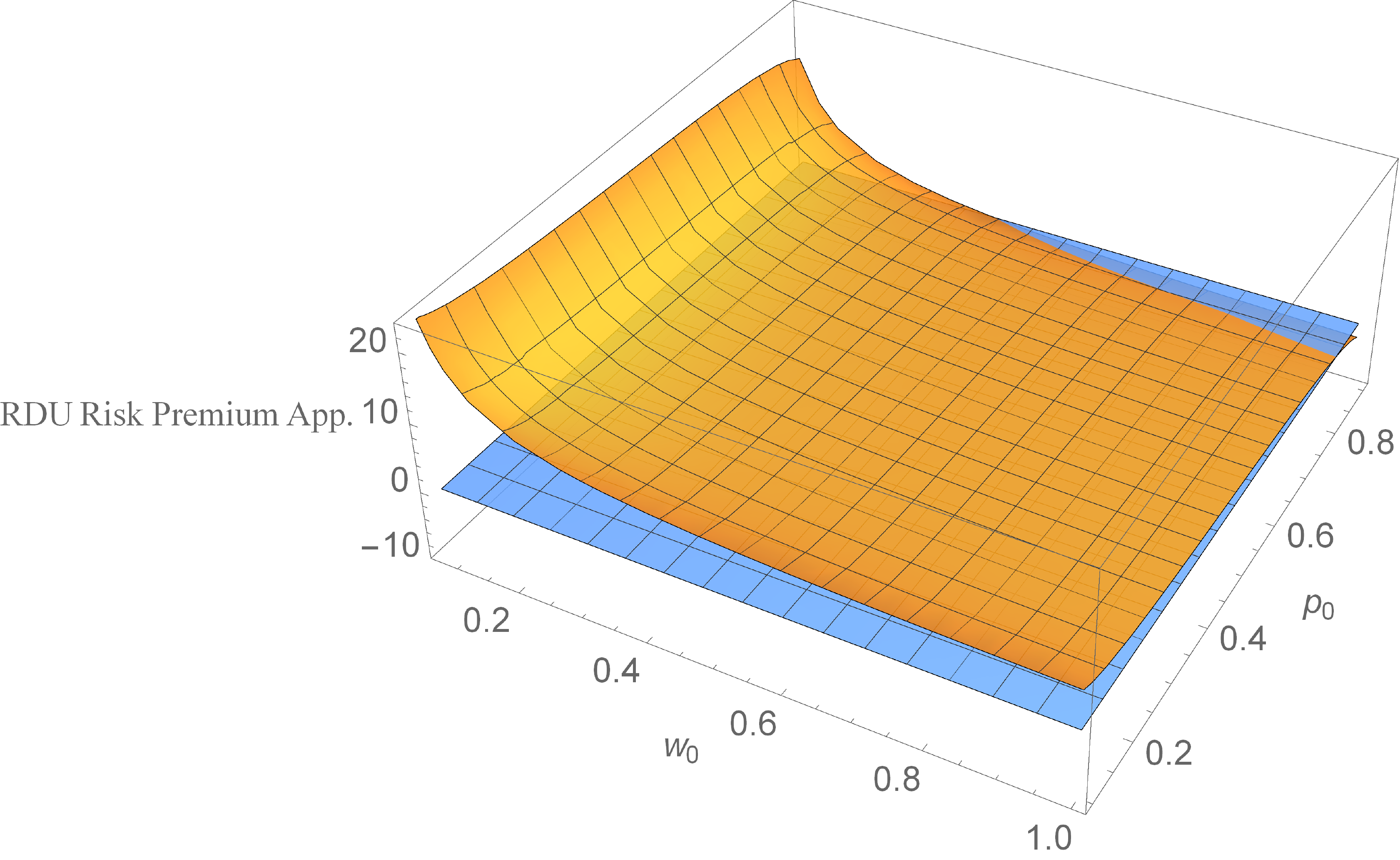}\vskip 0.1cm
\includegraphics[scale=0.43,angle=0]{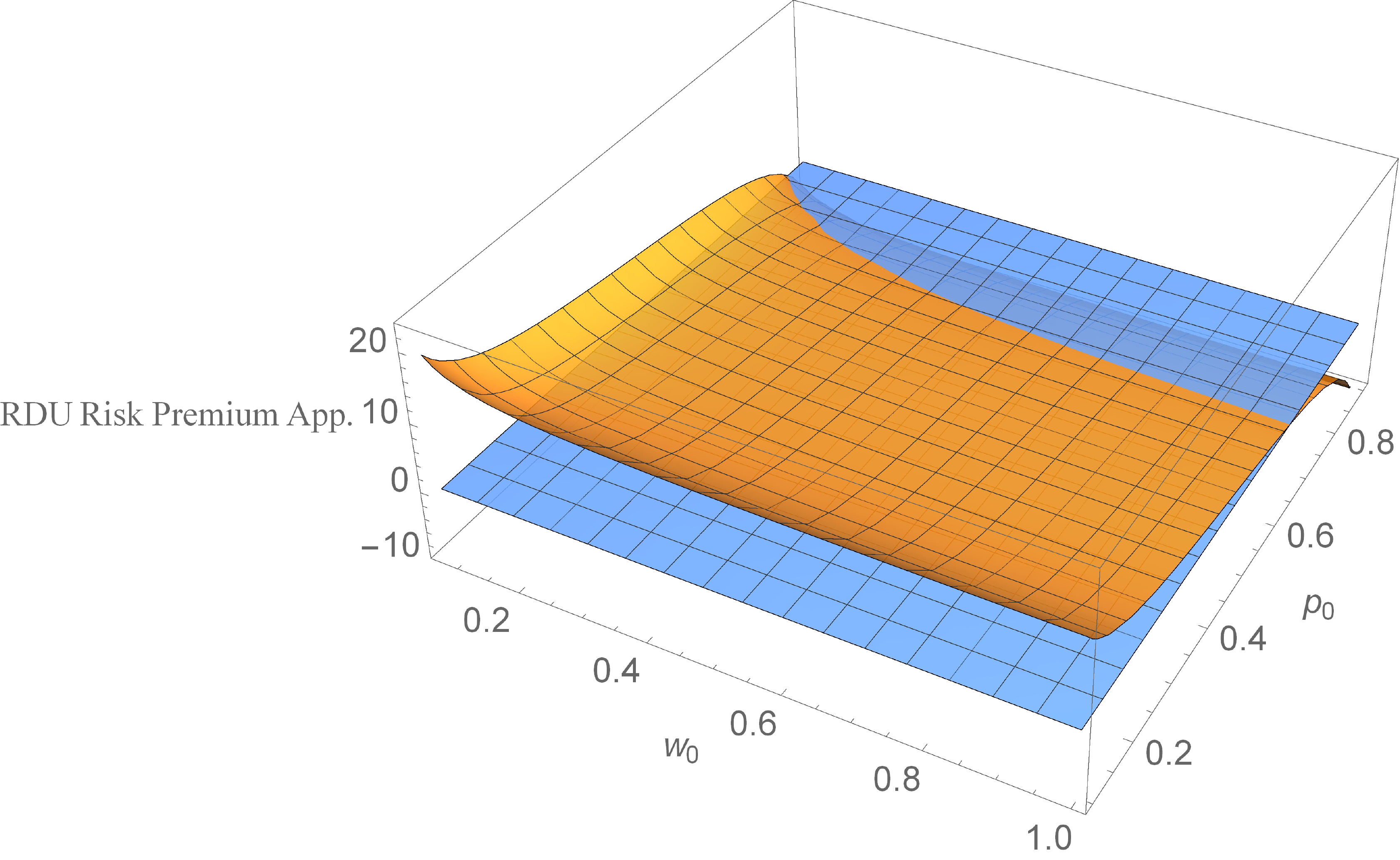}
\label{fig:RPvarmax}
\end{center}
\end{figure}

\end{document}